# Control of the effective value of the critical current of the RF SQUID by the high-frequency electromagnetic field

V.I. Shnyrkov[1,2], V.Yu. Lyakhno[2,3], O. A. Kalenyuk[1,2], D.G. Mindich[1], O.O. Leha[3], A.P. Shapovalov[1,2]

[1]Kiev Academic University, 36 Academician Vernadsky Boulevard, Kiev 03142, Ukraine

[2]G.V. Kurdyumov Institute for Metal Physics of NAS of Ukraine, 36 Academician Vernadsky Boulevard, 03142 Kyiv, Ukraine

[3]B.Verkin Institute for Low Temperature Physics and Engineering of NAS of Ukraine, 47 Nauky Ave., Kharkiv, 61103, Ukraine

E-mail: shapovalovap@gmail.com

An analysis of the influence of the high-frequency electromagnetic field on the amplitude-frequency and signal characteristics of RF SQUID and experimental verification are carried out. At low parameter $\beta_L$, the RF SQUID behavior is well described analytically by the theoretical model. In experiment, basic operation scheme is used in which the interferometer is inductively connected to a resonant tank circuit driven by RF current at a frequency close to the resonance frequency of the tank. It is shown that parameter $\beta_L$, which distinguishes between hysteretic and non-hysteretic regimes, can be effectively adjusted to a desired value by applying the high-frequency field of a certain amplitude and frequency much higher than tank resonant frequency. A significant increase in the conversion factor and sensitivity of the RF SQUID during this adjustment is discussed.

1. INTRODUCTION

The RF SQUID interferometer consists of a superconducting loop closed by a single Josephson junction, the critical current of which determines the characteristics and sensitivity of the parametric detector based on it. The well-developed technology for producing superconductor-insulator-superconductor (SIS) type contacts based on Nb layers [1] allows mass production of RF SQUIDs with the optimal value of the critical current for a given operating temperature. However, the critical currents of Josephson junctions composed of two-band superconductors and HTSCs for novel interferometers have usually a large spread and may be far from the optimal value.

The effect of the high-frequency (HF) field on the RF SQUID loop is quite different in the non-hysteretic $\beta_L = 2\pi L I_C / \Phi_0 < 1$ [2] and hysteretic $\beta_L > 1$ [3] regimes. Here $L$ is the geometric inductance of the interferometer, $I_C$ is the critical current of the Josephson junction and $\Phi_0 = \frac{h}{2e}$ is the magnetic flux quantum. This is associated with the distinctive difference in the potential energy of the interferometer in these two regimes (Fig. 1). In the resistive-capacitive shunted junction (RCSJ) model of the Josephson junction, it consists of the magnetic energy stored in the interferometer loop and the Josephson junction energy [4, 5]:

$$U(\varphi, \varphi_e) = \frac{I_c \Phi_0}{2\pi} \left[ \frac{(\varphi - \varphi_e)^2}{2\beta_L} - \cos\varphi \right] \quad (1)$$

where $\varphi = \frac{2\pi \Phi}{\Phi_0}$ is the phase difference at the Josephson junction, which is strictly related to the total magnetic flux piercing the loop $\Phi$, $\varphi_e = \frac{2\pi \Phi_e}{\Phi_0}$ is the normalized external magnetic flux $\Phi_e$. The interferometer behavior is reduced to the analysis of the motion of a particle with the mass $M = C \frac{\Phi_0}{2\pi}$ in one-dimensional potential (1), where $C$ is the junction capacity.

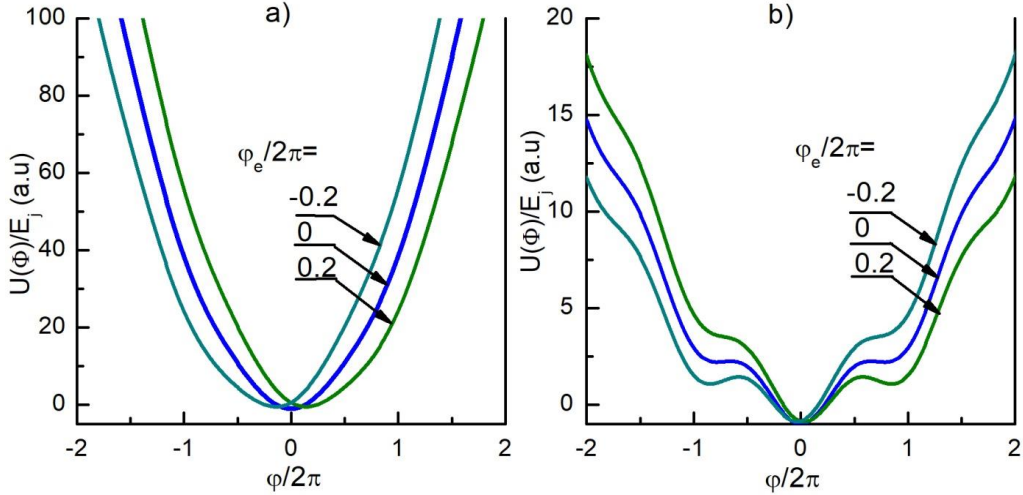

*Fig. 1*. Set of calculated potential energies of the RF SQUID as a function of $\varphi$, for three values of the external flux $\varphi_e / 2\pi = -0.2; 0; 0.2$: (a) in the non-hysteretic ($\beta_L = 0.5$) and (b) in the hysteretic ($\beta_L = 5.0$) regimes.

A rapid change in the potential energy with the frequency of the HF field will cause the particle to move slowly (with RF frequency) in a time-varying potential.

In the non-hysteretic regime $\beta_L < 1$, the potential energy has a single minimum (Fig. 1a), while the magnetic flux from the microwave field will induce small phase variations around this state. In the hysteretic regime, several metastable states separated by the potential barriers exist (Fig. 1b), while only one, with the lowest potential energy, is globally stable. Therefore, the microwave field will affect the probability of transition from one metastable state to another over the barriers that separate these metastable states, resulting in an increase in the slope of the steps of the IV curves in weak microwave fields [6].

## 2. MODEL CHARACTERISTICS OF RF SQUID

For simplification, let us assume that the influence of quantum and thermal $\gamma = 2\pi k_B T / I_C \Phi_0 \ll 1$ ($k_B$ is Boltzmann constant, $T$ is temperature) fluctuations on the interferometer are negligible. In the framework of the RCSJ model, we have equations that describe the behavior of the resonant tank with quality $Q >> 1$ and inductance $L_T$ coupled to the RF SQUID loop via the mutual inductance $M = k\sqrt{LL_T}$ ($k$ is coupling coefficient), in dimensionless quantities [4]:

$$\beta_L i = \varphi_e + \varphi_T - \varphi \qquad (2)$$

$$q\dot{\varphi} + \beta_L \sin\varphi + \varphi = \varphi_e + \varphi_T \qquad (3)$$

$$\ddot{\varphi}_T + Q^{-1}\dot{\varphi}_T + (1 - 2\zeta_0)\varphi_T = \varepsilon\cos\tau + k^2\beta_L \ddot{i} \qquad (4)$$

where $i = I/I_c$ is the circulating current in the interferometer, normalized to the critical current of the Josephson junction; $\varphi_T = \dfrac{2\pi\Phi_T}{\Phi_0}$ is the magnetic flux induced in the interferometer from the resonant circuit. Differentiation with respect to dimensionless time $\tau = \omega_T t$ is denoted by a dot ($\omega_T$ is circular tank resonance frequency). Parameter $q = \dfrac{\omega L}{R}$, where $R$ is normal resistance of the Josephson junction, characterizes the interferometer's delay relative to the pump frequency $\dfrac{\omega}{2\pi}$. The normalized detuning $\xi_0 = \dfrac{\omega - \omega_T}{\omega_T} \ll 1$ is small. $\varepsilon = (\dfrac{2\pi}{\Phi_0})\omega_T L_T I_P / \eta_0$ is the excitation amplitude, normalized to the flux-to-voltage conversion coefficient in the tank circuit $\eta_0 = (\omega/k)(L_T/L)^{1/2}$, and the coupling between the interferometer and the resonant circuit is considered weak $k^2 Q \approx 1$.

If the interferometer nonlinearity $\beta_L \ll 1$, and $k^2 \ll 1$, and $Q^{-1} \ll 1$ are small parameters, then the oscillations in the resonant parametric circuit can be considered quasi-harmonic, with amplitude $a(\tau)$ and phase $\theta(\tau)$ being slow functions of time $\tau$. Analytical expressions for the effective values of detuning $2\xi(a,\varphi_e)$ and damping $2\delta(a,\varphi_e)$ of the resonant tank, keeping linear terms by $\beta_L$, are given by [4]:

$$2\xi(a,\varphi_e) = 2\xi_0 + \frac{k^2}{1+q^2}[-q^2 - \beta_L \frac{2J_1(a)}{a}\frac{1-q^2}{1+q^2}\cos\varphi_e] \qquad (5),$$

$$2\delta(a,\varphi_e) = Q^{-1} + \frac{k^2}{1+q^2}[q - \beta_L \frac{2J_1(a)}{a}\frac{2q}{1+q^2}\cos\varphi_e] \qquad (6),$$

where $J_1(a)$ is the first-order Bessel function of the first kind.

Equations (5, 6) describe the behaviour of the resonant tank circuit with nonlinear reactive characteristics introduced by the interferometer. Its resonance curve is symmetrical (at least at small $\beta_L$) about the line bent according to equation (5). At $|\cos\varphi_e| = 0$, this curve is becomes a straight line like in a usual linear tank circuit, while at $|\cos\varphi_e|=1$, it has the maximum curvature. At sufficiently large values of the parameter

$$k^2 Q \beta_L > 1 \qquad (7)$$

the amplitude-frequency characteristic (AFC) in the non-hysteretic regime can be multivalued and have "breaking points" with vertical tangents (Fig. 2), in the vicinity of which the value of the conversion coefficients sharply increases (formally to infinity) [2, 4, 7].

$$\eta = \left|\frac{da}{d\varphi_e}\right| \gg \eta_0 \qquad (8)$$

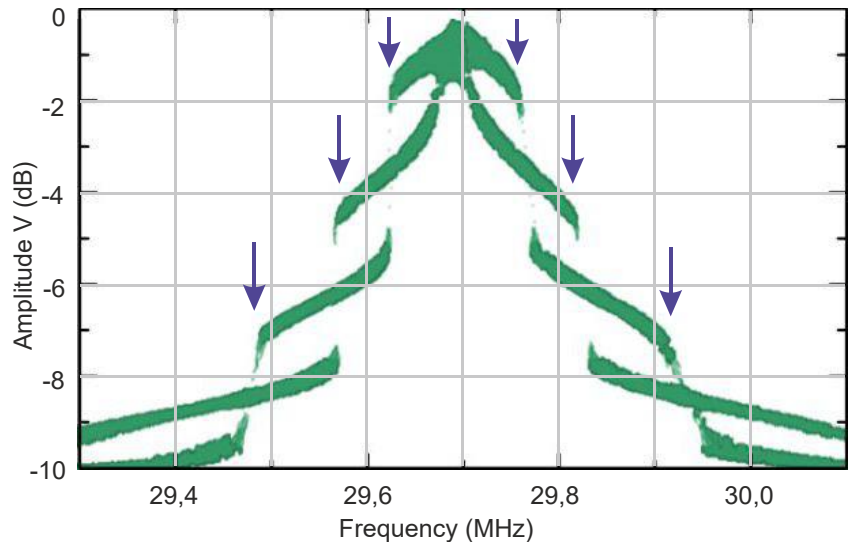

*Fig. 2.* AFCs of the RF SQUID in the non-hysteretic regime $k^2 Q \beta_L \gg 1$ for two values of the external flux $\varphi_e \approx \pi$ and $\varphi_e \approx 2\pi$ at a constant RF pumping amplitude. The "breaking points" (indicated by arrows) are symmetrically located at negative and positive detunings, as the parameter $\beta_L \approx 0.1 \ll 1$.

If the critical current of the Josephson junction is large enough, the interferometer will operate in the hysteretic regime. For $k^2 Q \gtrsim 1$, condition (7) can be satisfied by raising the temperature so that the RF SQUID changes to a non-hysteretic regime $\beta_L(T) = 2\pi L I_C(T)/\Phi_0 < 1$. However, we will consider the temperature to be constant from now on and examine the characteristics of the RF SQUID, in which a high-frequency (microwave) field $\tilde{\varphi}\sin\nu t$ is introduced, where $\tilde{\varphi} = \frac{2\pi}{\Phi_0}\tilde{\Phi}$ represents the normalized amplitude of the microwave magnetic flux.

3. THE EFFECT OF HIGH-FREQUENCY ELECTROMAGNETIC FIELD ON THE CHARACTERISTICS OF RF SQUID

It is worth noting that due to the significant inequality $\nu \gg \omega_T$, the resonant circuit "sees" the averaged (over high frequency) value of potential energy over each RF period. Keeping the condition $\beta_L \ll 1$ unchanged, let us add an HF flux $\tilde{\varphi}\sin\nu t$ with frequency $\nu \gg \omega_T/2\pi$ to the external, time-independent flux $\varphi_e$. It is seen from the inequality that HF corresponds practically to the microwave frequency range. For $\beta_L \ll 1$, in the zero approximation for $\beta_L$, we have

$$\varphi^{(0)} = \varphi_e + \frac{a}{1+q^2}[\cos(\tau+\Theta) + q\sin(\tau+\Theta)] + \frac{\tilde{\varphi}}{(1+\nu q)^2}[\cos\nu\tau + \nu q\sin\nu\tau]. \quad (9)$$

When finding the next approximation, we neglect the resulting combination frequencies, which can contribute to the first harmonic, but it will be small due to the large frequency separation. As a result, we obtain a system of equations for the effective values of detuning and damping of the resonant circuit depending on the amplitude of the high-frequency field [7]:

$$2\xi(a,\varphi_e,\tilde{\varphi}) = 2\xi_0 + \frac{k^2}{1+q^2}[-q^2 - \beta_L J_0(z)\frac{2J_1(a)}{a}\frac{1-q^2}{1+q^2}\cos\varphi_e] \quad (10),$$

$$2\delta(a,\varphi_e,\tilde{\varphi}) = Q^{-1} + \frac{k^2}{1+q^2}[q - \beta_L J_0(z)\frac{2J_1(a)}{a}\frac{2q}{1+q^2}\cos\varphi_e] \quad (11),$$

here $J_0(z)$ is the Bessel function with argument $z = \tilde{\varphi}/[1+(\nu q)^2]^{1/2}$.

Comparing equations (5, 6) with (10, 11), we can observe that, at a temperature $T = const$, the presence of the HF field formally renormalizes the value of the main parameter $\beta_L$ of the RF SQUID, while maintaining dependencies on $q$, on the amplitude of the resonant circuit pumping, and on the low-frequency (signal) flux $\varphi_e$ unchanged. In this case, the important inequality (7) can also be satisfied for excitation amplitudes $\varepsilon \gtrsim Q^{-1}$ by the replacement $\beta_L \to \beta_L J_0(z)$. At large values of the HF electromagnetic field $|\tilde{\varphi}| \gg 1$, equations (10, 11) describe a linear resonant circuit.

Despite the fact that expressions (5, 6) and (10, 11) are derived for small $\beta_L \ll 1$, they can practically be used up to $\beta_L \lesssim 0.3$, and the general behavior of the RF SQUID in the HF electromagnetic field will qualitatively remain practically unchanged up to $\beta_L \approx 1$.

In the hysteretic regime, the situation is more complex, and analytical formulas are unavailable. Results from numerical simulations [6] indicate that weak HF fields will increase the slope of the "steps" in the hysteretic regime and lead to a decrease in sensitivity. Essentially, this is associated with an expansion of the probability density of phase jumps in the interferometer [3], resulting in additional losses. The effect of strong HF electromagnetic fields can be relatively easily revealed experimentally.

4. EXPERIMENTAL RESULTS AND MEASUREMENTS

The experiments were performed at a fixed temperature $T \cong 4.2$ K. The RF SQUID was placed in a superconducting lead shield to protect it against external fields. The tank circuit with quality factor $Q \approx 200$ had resonance frequency $\omega_T/2\pi = 1/2\pi(L_T C_T)^{1/2} \approx 30$ MHz, where $L_T \approx 3 \times 10^{-7}$ H and $k^2 Q \approx 1.5$. The RF pumping of the tank circuit at frequencies close to the resonance frequency was performed using a standard frequency-sweep generator. To reduce the thermal noise of the coaxial cable, the first amplifier stage was placed in liquid helium. The HF field $\tilde{\varphi}\sin\nu t$ with frequency $\nu = 1$ GHz from the generator was put into the RF SQUID through an attenuator via a separate coaxial cable weakly coupled to the interferometer.

The geometric inductance of the interferometer $L \approx 3 \times 10^{-10}$ H was chosen to be much less than so called fluctuation inductance $L_F$, $L \ll L_F = (\Phi_0/2\pi)^2/k_B T$, at $T \cong 4.2$ K in order to the noise flux associated with thermal fluctuations could not break the coherence in the interferometer loop. Estimating the parameters $\beta_C = 2\pi R^2 I_C C/\Phi_0$ and $q$ gave corresponding values of $\approx 4 \times 10^{-2}$ and $\approx 10^{-2}$.

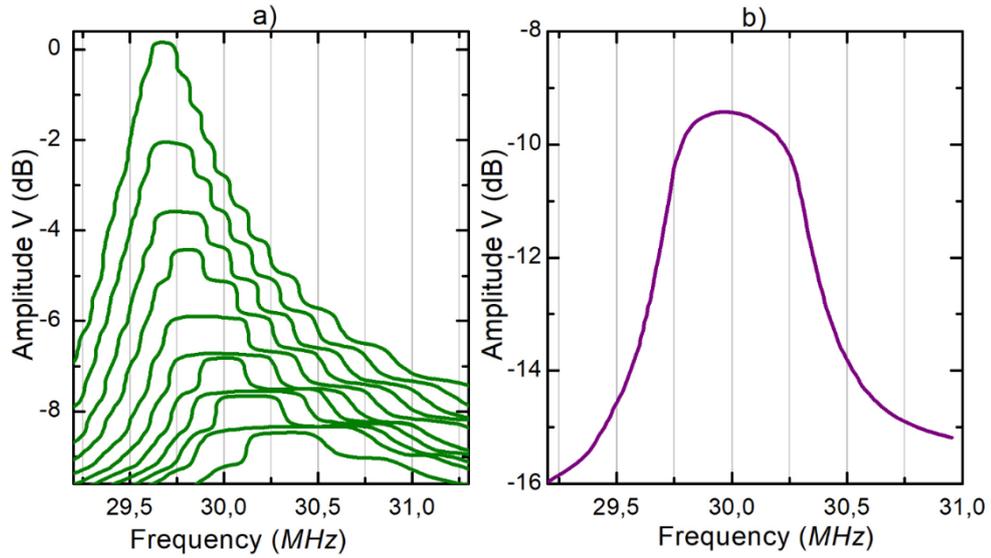

*Fig. 3*. Set of AFCs of the RF SQUID in the hysteretic regime $\beta_L \approx 10$ without HF electromagnetic field at temperature 4.2 K: (a) The value of external magnetic flux $\Phi_e \approx \Phi_0$, and pumping amplitude changes every 1 dB (b) AFC of the RF SQUID at the pumping current amplitude $I_P$ corresponding to the first step at $\Phi_e \approx \Phi_0/2$.

In the absence of the HF field (Fig. 3a), AFCs of the RF SQUID were obtained at temperature 4.2 K, typical for the hysteretic regime ($\beta_L \approx 10$). Then the external flux was fixed to observe only the first step, as shown in Figure 3(b).

After that, an HF field with a frequency of 1 GHz was introduced into the interferometer, which, at small amplitudes, resulted in an additional slope on the step (Fig. 4), consistent with the findings of [6]. If the amplitude further increases, it rapidly modifies the AFC as if reducing the junction critical current and, consequently, the effective parameter $\beta_L$ down to values $\beta_L \approx 1$.

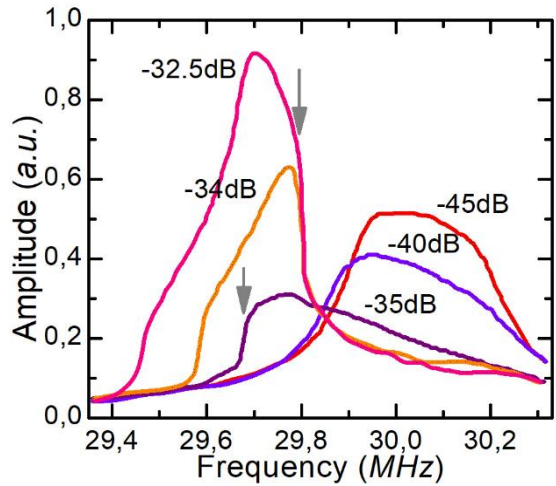

*Fig. 4*. Set of AFCs of the RF SQUID in HF field at a frequency of $\nu = 1$ GHz. The curves parameter is the HF amplitude of electromagnetic field (dB). The RF pumping current $I_P$ and constant magnetic flux are fixed for the entire curve set. The regions with steeper slopes are indicated by arrows.

Starting from the AFC obtained at -35 dB, the curves typical for the non-hysteretic regime are observed (curves -35 dB, -34 dB, -32.5 dB) with quasi-periodic variations of the tank resonance frequency as a function of HF power. It is worth noting that the regions on these AFCs resembling

"breaking points" with nearly vertical tangents, specific for the non-hysteretic regime with valid condition $k^2 Q \beta_L > 1$, are clearly observed (Fig. 4).

If the operating point of the RF SQUID is located in the region with large derivative $\left|\frac{da}{d\varphi_e}\right|$, then the small-signal conversion coefficient $\eta$ becomes 5-6 times higher than in the standard hysteretic regime $\eta_0 = (\omega/k)(L_T/L)^{1/2}$. Such an increase in the conversion coefficient significantly reduces the contribution of the amplifier $\delta\varepsilon_A \sim \frac{\langle V_{NA} \rangle^2}{|\eta|^2}$ to the RF SQUID noise, thereby improving its sensitivity.

## 5. CONCLUSIONS

The literature mostly discussed [4, 6] the increase in the slope of the steps $\alpha_{JB}$ in RF SQUIDs with $\beta_L > 1$ under the influence of weak microwave fields, and the associated degradation of the device sensitivity [5]. The phenomenological parameter $\alpha_{JB}$, which describes the junction noise, is associated with the change in the probability of jumps between metastable states [7]. It typically depends on the amplitude of the HF (microwave) field, as seen clearly in Fig. 4.

However, the experiments made in this work demonstrate that variations of the HF field amplitude effectively transform the characteristics of RF SQUIDs from a classical hysteretic regime to a regime formally similar to non-hysteretic. Such control of the main parameter of the RF SQUID is particularly interesting for two-zone superconductors [10, 11, 12] and HTSs [13,14]. Somewhat unexpectedly, the regions with conversion coefficients exceeding the classical value for the hysteretic regime were observed at the AFCs of RF SQUIDs during this transformation. This feature allows one to reduce essentially the contribution of the amplifier intrinsic noise and improve the device sensitivity as a whole [14]. Undoubtedly, this property of the nonlinear resonance of the RF SQUID in HF electromagnetic field is of great interest for the development of practical devices and requires further investigation.

ACKNOWLEDGEMENTS

We grateful Oleg Turutanov for useful discussions and technical help. This work was carried out within the framework of the project G5796 funded by the NATO SPS Program jointly with MES of Ukraine and supported by the IEEE program "Magnetism for Ukraine 2023", project number 9918.

REFERENCES

1. *Niobium Integrated Circuit Fabrication, Process #03-10-45, Design rules, Revision #25*, 12/12/2012, Hypres, Inc. (2012). Available at https://www.hypres.com/wp-content/uploads/2010/11/DesignRules-4.pdf.
2. I.M. Dmitrenko, G.M. Tsoi, V.I. Shnyrkov, V.V. Kartsovnik, *J Low Temp. Phys.* **49**, 417 (1982). https://doi.org/10.1007/BF00681894;
3. V.I. Shnyrkov, A.P. Shapovalov, V.Yu. Lyakhno, A.O. Dumik, A.A. Kalenyuk, *Superconductor Science Technology*, **36(3)**, 035005 (2023). https://doi.org/10.1088/1361-6668/acb10e
4. K.K. Likharev, Dynamics of Josephson Junctions and Circuits, Routledge (1986). https://doi.org/10.1201/9781315141572
5. The SQUID Handbook: Fundamentals and Technology of SQUIDs and SQUID Systems, J. Clarke, A. I. Braginski (eds.), Wiley-VCH, Weinheim (2004). https://doi.org/10.1002/3527603646
6. L.D. Jackel, R.A. Buhrman, *J. Low Temp. Phys*. **19**, 201 (1975). https://doi.org/10.1007/BF00116178
7. O.G. Turutanov, V.Yu. Lyakhno, M.E. Pivovar, V.I. Shnyrkov, *Low Temperature Physics*, **45**, 60. (2019) https://doi.org/10.1063/1.5082311
8. V.I. Shnyrkov, G.M. Tsoi, *Principles and Applications of Superconducting Quantum Interference Device*s, (Ed. A. Barone), (Singapore: World Scientific:1992), pp. 77. https://doi.org/10.1142/9789814439718_0002
9. V.I. Shnyrkov, V.A. Khlus, and G.M. Tsoi, *J. Low Temp. Phys.,* **39**, 477 (1980).
10. Y. S. Yerin & A. N. Omelyanchouk, *Low Temp. Phys.,* **969** (2010). https://doi.org/10.1063/1.3518605
11. V Tarenkov, A Shapovalov, O Boliasova, M Belogolovskii, A Kordyuk, *Low Temp. Phys.,* 47 (2), 101-105 (2021) https://doi.org/10.1063/10.0003168
12. AA Kalenyuk, AL Kasatkin, SI Futimsky, AO Pokusinskiy, TA Prikhna, AP Shapovalov, VE Shaternik, Sh Akhmadaliev, *Superconductor Science and Technology* 36 (3), 035009 (2023) https://doi.org/10.1088/1361-6668/acb110
13. Fundamentals and Frontiers of the Josephson Effect, F. Tafuri(ed), Springer, Cham (2019) https://doi.org/10.1007/978-3-030-20726-7.
14. D. Koelle, R. Kleiner, F. Ludwig, E. Dantsker* and John Clarke High-transition-temperature superconducting quantum interference devices, *Reviews of Modern Physics*, **71**, 631-686 (1999) https://doi.org/10.1103/RevModPhys.71.1249